\documentclass[12pt,a4paper]{article}
\usepackage{graphicx}
\begin{document}

\title{Asymmetry of in-medium $\rho $-mesons\\ 
as a signature of Cherenkov effects}

\author{I.M. Dremin\footnote{email: dremin@lpi.ru},  V.A. Nechitailo\footnote{email: nechit@lpi.ru}
\\
{\it P.N.Lebedev Physics Institute RAS, 119991 Moscow, Russia}}
 
\maketitle

\begin{abstract}
Cherenkov gluons may be responsible for the asymmetry of dilepton mass
spectra near $\rho $-meson observed in experiment. They can be produced only 
in the low-mass wing of the resonance. Therefore the dilepton mass spectra
are flattened there and their peak is slightly shifted to lower masses compared 
with the in-vacuum $\rho $-meson mass. This feature must be common for all 
resonances.
\end{abstract}


There exist numerous experimental data \cite{1, 2, 3, 4, 5, Muto, 6, 7, 8} about
the in-medium modification of widths and positions of prominent vector-meson
resonances. They are mostly obtained from the shapes of dilepton mass and
transverse momentum spectra in nucleus-nucleus collisions. Such in-medium
effects were tied theoretically to chiral symmetry restoration a long
time ago \cite{9}.

The dilepton mass spectra decrease approximately exponentially with 
increase of masses albeit with substantial declines from the average 
approximation of the general trend by the exponent in the 
low-mass region. A significant excess of low-mass dilepton pairs yield over 
expectations from hadronic decays is observed in experiment. The shape of the 
excess mass spectra shown in \cite{1, 2, 3} is dominated by $\rho $-mesons. 
Their ratio to other vector 
meson resonances can be estimated as $\rho :\omega :\phi $=10:1:2.

Several approaches have been advocated  for explanation of the excess.
Strong dependence of the parameters of the effective Lagrangian on the 
temperature and the chemical potential was assumed in \cite{10, 11}. The 
hydrodynamical evolution was incorporated in \cite{12} to describe the spectra.
The QCD sum rules and dispersion relations have been used \cite{13, 14} to
show that condensates decrease in the medium leads to both broadening and 
slight downward mass shift of resonances. The similar conclusions have been 
obtained from more traditional attempts using either the empirical scattering 
amplitudes with parton-hadron duality \cite{15, 15a} or the hadronic many-body 
theory \cite{16, 17, 18}.

In the latest approach, which pretends on the best description of experimental
plots, the in-medium V-meson spectral functions are evaluated.
The excess of dilepton pairs below $\rho $-mass is ascribed to anti-/baryonic
effects. This conclusion is the alternative to more 
common ideas about the chiral restoration at high energies. It asks 
for some empirical constraints to fit the observed excess.

In this paper we propose another possible source of low-mass lepton
pairs. Namely, the emission of Cherenkov gluons may provide a substantial 
contribution to the low mass region. 

Considered first for processes at very high energies \cite{19}, the idea about 
Cherenkov gluons was extended to resonance production \cite{20, 21a}. For 
Cherenkov effects to be pronounced in ordinary or nuclear matter, the (either 
electromagnetic or nuclear) index of refraction of the medium $n$ should be 
larger than 1. Qualitatively, the observed low mass excess of lepton pairs 
is easy to ascribe to the gluonic Cherenkov effect if one reminds that 
the index of refraction of any medium exceeds 1 within the lower wing of any 
resonance (the $\rho $-meson, in particular). 

This feature is well known in electrodynamics (see, e.g., Fig. 31-5 in 
\cite{21}) where the atoms behaving as oscillators emit as Breit-Wigner
resonances when get excited. This results in the indices of refraction larger 
than 1 within their low-energy wings. In QCD, one can imagine that the nuclear 
index of refraction for gluons in the hadronic medium behaves in a similar way 
in the resonance regions. This statement is more general and can be valid also 
at other energies if the relation (see, e.g., \cite{22}) between the index of 
refraction and the forward scattering amplitude $F(E, 0^o)$ is fulfilled not 
only for photons but for gluons as well:
\begin{equation}
\Delta n={\rm Re}n-1\propto {\rm Re}F(E, 0^o)/E.    \label{delt}
\end{equation}
Here $E$ is the photon (gluon) energy. In classical electrodynamics, it is
the dipole excitation of atoms in the medium by light which results in the
Breit-Wigner shape of the amplitude $F(E, 0^o)$. In hadronic medium, there 
should be some modes (quarks, gluons or their preconfined bound states, 
condensates, blobs of hot matter...?) which can get excited by the impinging 
parton, radiate coherently if $n>1$ and hadronize at the final stage as 
hadronic resonances \cite{20, 21a}. The hadronic Cherenkov effect can provide
insight into the substructure of the medium formed in nucleus-nucleus
collisions. The resonance amplitudes are chosen for $F(E, 0^o)$ at 
comparatively low energies.

The scenario, we have in mind, is as follows. The initial parton, belonging
to a colliding nucleus, emits a gluon which traverses the nuclear medium.
On its way, it collides with some internal modes. Therefore it affects the 
medium as an "effective" wave which accounts also for the waves emitted by 
other scattering centers (see, e.g., \cite{22}). Beside incoherent scattering,
there are processes which can be described as the refraction of the initial 
wave along the path of the coherent wave. The Cherenkov effect is the induced 
coherent radiation by a set of scattering centers placed on the way of 
propagation of a gluon. That is why the forward scattering amplitude plays 
such a crucial role in formation of the index of refraction. At low energies 
its excess over 1 is related to the resonance peaks as dictated by the
Breit-Wigner shapes of the amplitudes. In experiment, usual resonances are 
formed during the color neutralization process. However, only those gluons 
whose energies are within the left-wing resonance region of $n>1$ give rise 
also to collective Cherenkov effect proportional to $\Delta n$.

Thus, apart from the ordinary Breit-Wigner shape of the cross section for 
resonance production, the dilepton mass spectrum would acquire the additional
term proportional to $\Delta n$ at masses below the resonance peak. 
Therefore its excess near the $\rho $-meson can be described by the following 
formula\footnote{We consider only $\rho $-mesons here. To include other mesons,
one should evaluate the corresponding sum of similar expressions.}
\begin{equation}
\frac{dN_{ll}}{dM}=\frac {A}{(m_{\rho }^2-M^2)^2+M^2\Gamma ^2}
\left(1+w\frac{m_{\rho }^2-M^2}{M^2 }\theta (m_{\rho}-M)\right)    \label{ll}
\end{equation}
Here $M$ is the total c.m.s. energy of two colliding objects (the dilepton
mass), $m_{\rho }$=775 MeV is the in-vacuum $\rho $-meson mass.
The first term corresponds to the Breit-Wigner cross section. According to the
optical theorem it is proportional to the imaginary part of the forward
scattering amplitude. The second term is proportional to $\Delta n$ where
it is taken into account that the ratio of real to imaginary parts of
Breit-Wigner amplitudes is
\begin{equation}
\frac {{\rm Re} F(M, 0^o)}{{\rm Im} F(M, 0^o)}=\frac {m_{\rho }^2-M^2}{M\Gamma }.
\label{reim}
\end{equation}
This term vanishes for $M>m_{\rho }$ because only positive $\Delta n$ lead
to the Cherenkov effect. Namely it describes the distribution of
masses of Cherenkov states. In these formulas, one should take into account 
the in-medium modification of the height of the peak and its width. In 
principle, one could consider $m_{\rho }$ as a free in-medium parameter as well.
We rely on experimental findings that its shift in the medium is small.
All this asks for some dynamics to be known. In our approach, it is
not determined. Therefore, first of all, we just fit the parameters $A$ 
and $\Gamma $ by describing the shape of the mass spectrum at $0.75<M<0.9$ GeV 
measured in \cite{3} and shown in Fig. 1. In this way we avoid any
strong influence of the $\phi $-meson. Let us note that 
$w$ is not used in this procedure. The values $A$=104 GeV$^3$ 
and $\Gamma =0.354$ GeV were obtained. The width of the in-medium 
peak is larger than the in-vacuum $\rho $-meson width equal to 150 MeV.

Thus the low mass spectrum at $M<m_{\rho }$ depends only on a single parameter
$w$ which is determined by the relative role of Cherenkov effects and 
ordinary mechanism of resonance production.  It is clearly seen from 
Eq. (\ref{ll}) that the role  of the second term in the brackets increases 
for smaller masses $M$. The excess spectrum in the mass region from 0.4 GeV 
to 0.75 GeV has been fitted by $w=0.19$. The slight downward shift 
about 40 MeV of the peak of the distribution compared with $m_{\rho }$ may
be estimated from Eq. (\ref{ll}) at these values of the parameters. This 
agrees with the above statement about small shift compared to $m_{\rho }$.
The total mass spectrum (the dashed line) and its widened Breit-Wigner 
component (the solid line) according to Eq. (\ref{ll}) with the chosen 
parameters are shown in Fig. 1. The overall description of experimental 
points seems quite satisfactory. The contribution of Cherenkov gluons (the 
excess of the dashed line over the solid one) constitutes the noticeable part 
at low masses. The formula (\ref{ll}) must be valid in the vicinity of the 
resonance peak. Thus we use it for masses larger than 0.4 GeV only.
\begin{figure}[ht]
\includegraphics[width=\textwidth]{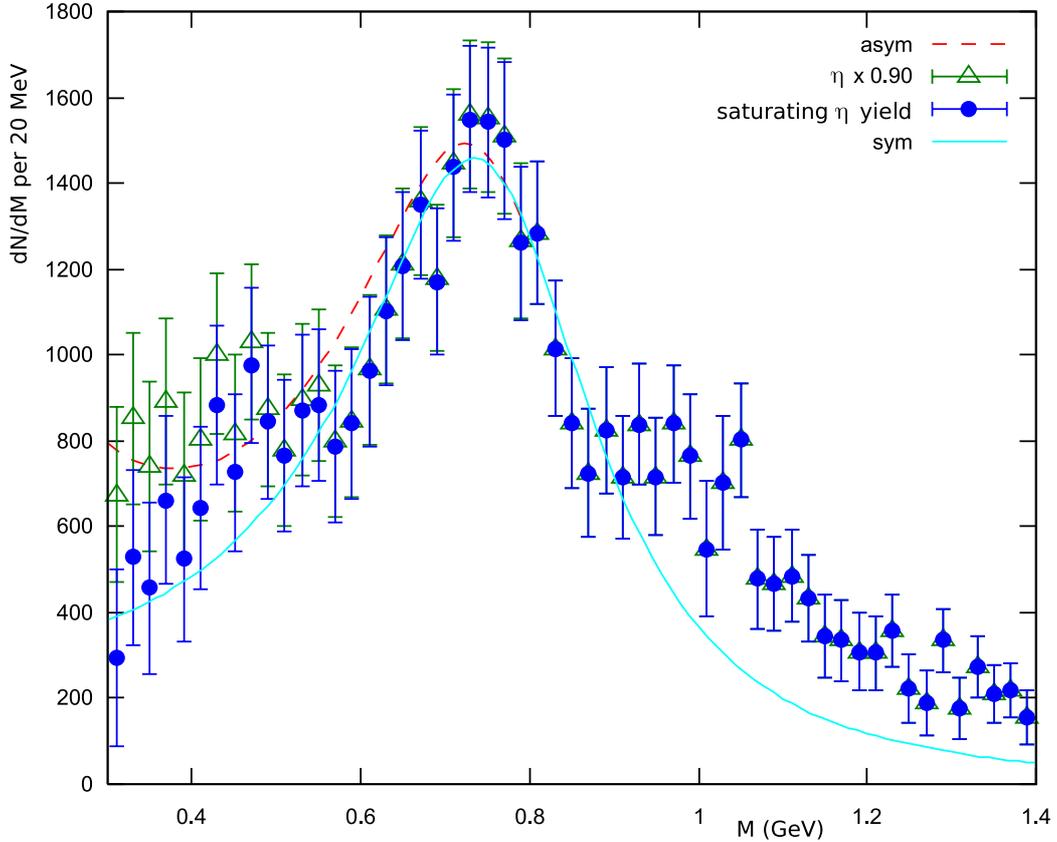}
 \caption{Excess dilepton mass spectrum in semi-central In(158 AGeV)-In of NA60 (dots) 
compared to the in-medium $\rho $-meson peak with additional Cherenkov effect 
(dashed line).}
\end{figure}

The experimental data plotted in Fig. 1 have not been corrected for
 the acceptance of the experiment, which strongly depends on mass and
 transverse momentum of the muon pairs. However, due to an approximate
 cancellation between the variations of the thermal radiation mediated
 by the rho and those of the acceptance, the data as shown can roughly
 be interpreted as spectral function of the rho, averaged over momenta
 and the complete space-time evolution of nuclear collision \cite{3}. To use
 these data without further corrections is therefore justified as long
 as the $p_T$  spectra of the radiation and those of the Cherenkov process
 are not dramatically different. From general principles one would expect 
slightly lower $p_T$ for low-mass dilepton pairs from coherent Cherenkov 
processes than for incoherent scattering. Qualitatively, this conclusion 
is supported by experiment \cite{3}. The
Cherenkov dominance region of masses from 400 MeV to 600 MeV below the 
$\rho $-resonance has softer $p_T$-distribution compared to the resonance 
region from 600 MeV to 900 MeV filled in by usual incoherent scattering.
More accurate statements can be obtained after the microscopic theory
of Cherenkov gluons developed. 

We should mention that the expression (\ref{ll}) may be applied for 
$\Delta n \ll 1$. The RHIC experiments revealed rather large 
$\Delta n \approx 2$. If the same values are typical at lower energies 
of SPS then the more general formulas (see \cite{21a}) should be used. 
The qualitative conclusions stay valid.

Whether the in-medium Cherenkov gluonic effect is as strong as shown in 
Fig. 1 can be verified by measuring the angular distribution of the lepton
pairs with different masses. The trigger-jet experiments similar to that at 
RHIC are necessary to check this prediction. One should measure the
angles between the companion jet axis and the total momentum of the lepton
pair. The Cherenkov pairs with masses between 0.4 GeV and 0.7 GeV
should tend to fill in the rings around the jet axis. The angular radius 
$\theta $ of the ring is determined by the usual condition 
\begin{equation}
\cos \theta = \frac {1}{n}        \label{thet}
\end{equation}
as discussed in more detail in \cite{20}. 

Another way to demonstrate it is to measure the average mass of lepton pairs
as a function of their polar emission angle (pseudorapidity) with the
companion jet direction chosen as $z$-axis. Some excess of low-mass pairs 
may be observed at the angle (\ref{thet}). 
Baryon-antibaryon effects can not possess signatures similar to these ones.

In practice, these procedures can be quite complicated at comparatively low 
energies if the momenta of decay products are comparable to the transverse 
momentum of the resonance. It can be a hard task to pair leptons in reliable 
combinations. The Monte Carlo models could be of some help.

In non-trigger experiments like that of NA60 there is another obstacle. 
Everything is averaged over directions of initial  partons. Different partons 
are moving in different  directions. The angle $\theta $, measured from the
direction of their initial momenta, is the same but the total angles are 
different, correspondingly. The averaging procedure would shift the maxima 
and give rise to more smooth distribution. Nevertheless, some indications 
on the substructure with maxima at definite angles have been found at the 
same energies by CERES collaboration \cite{26}. It is not clear yet 
if it can be ascribed to Cherenkov gluons. To recover a definite maximum,
it would be better to detect a single parton jet, i.e. to have a trigger. 

The prediction of asymmetrical in-medium widening of {\bf any} resonance at its
low-mass side due to Cherenkov gluons is universal. This universality is 
definitely supported by experiment. Very clear signals of the excess on 
the low-mass sides of $\rho $, $\omega$ and $\phi$ mesons have been seen in 
\cite{5, Muto}. This effect for $\omega $-meson is also studied in
\cite{7}. Slight asymmetry of $\phi $-meson near 0.9 - 1 GeV is noticeable in
the Fig. 1 shown above but the error bars are large there. We did not try to 
fit it just to deal with as small number of parameters as possible. There are 
some indications at RHIC (see Fig. 6 in \cite{6}) on this effect for 
$J/\psi $-meson.

At much higher energies one can expect better alignement of the momenta of
initial partons. This would favour the direct 
observation of emitted by them rings in non-trigger experiments. The first 
cosmic ray event \cite{23} with ring structure gives some hope that at LHC 
energies the initial partons are really more aligned and this effect can be 
found. The possible additional signature at high energies could be the 
enlarged transverse momenta of particles within the ring.

To conclude, the new mechanism is proposed for explanation of the low-mass 
excess of dilepton pairs observed in experiment. It is the Cherenkov gluon 
radiation which adds to the ordinary processes at the left wing of any 
resonance.

\bigskip

{\bf\large Acknowledgments }\\

We thank S. Damjanovic for providing us with experimental data. I.D. is
grateful to S. Damjanovic and H. Specht for very illuminating and fruitful 
discussions, in particular, on the role of the experimental acceptance.

This work has been supported in part by the RFBR grants 06-02-16864, 06-02-17051.


\begin{thebibliography}{99}

\bibitem{1}
G. Agakichiev et al. (CERES), Phys. Rev. Lett. {\bf 75}, 1272 (1995); Phys. Lett. 
{\bf B422}, 405 (1998); Eur. Phys. J. {\bf C41}, 475 (2005). 
\bibitem{2}
D. Adamova et al. (CERES), Phys. Rev Lett. {\bf 91}, 042301 (2003); {\bf 96},
152301 (2006); nucl-ex/0611022.
\bibitem{3}
R. Arnaldi et al. (NA60), Phys. Rev. Lett. {\bf 96}, 162302 (2006);  
S. Damjanovic et al. (NA60), Eur. Phys. J. {\bf C49}, 235 (2007), 
nucl-ex/0609026 and Nucl. Phys. {\bf A783}, 327 (2007), nucl-ex/0701015.
\bibitem{4}
D. Trnka et al., Phys. Rev. Lett. {\bf 94}, 192303 (2005). 
\bibitem{5}
M. Naruki et al. (KEK), Phys. Rev. Lett. {\bf 96}, 092301 (2006), nucl-ex/0504016.
\bibitem{Muto}  
R. Muto et al. (KEK), Phys. Rev. Lett. {\bf 98}, 042501 (2007), nucl-ex/0511019.
\bibitem{6}
A. Kozlov (PHENIX), nucl-ex/0611025.
\bibitem{7}
M. Kotulla (CBELSA/TAPS), nucl-ex/0609012.
\bibitem{8}
I. Tserruya, Nucl. Phys. {\bf A774}, 415 (2006).
\bibitem{9}
R. Pisarski, Phys. Lett. {\bf B110}, 155 (1982).
\bibitem{10}
M. Harada and K. Yamawaki, Phys. Rep. {\bf 381}, 1 (2003).
\bibitem{11}
G.E. Brown and M. Rho, Phys. Rev. Lett. {\bf 66}, 2720 (1991); Phys. Rep.
{\bf 269}, 333 (1996); Phys. Rep. {\bf 363}, 85 (2002).
\bibitem{12}
K. Dusling, D. Teaney and I. Zahed, nucl-th/0604071.
\bibitem{13}
S. Leupold, W. Peters and U. Mosel, Nucl. Phys. {\bf A628}, 311 (1998).
\bibitem{14}
J. Ruppert, T. Renk and B. Muller, Phys. Rev. {\bf C73}, 034907 (2006); 
hep-ph/0612113.
\bibitem{15}
V.L. Eletsky, M. Belkacem, P.J. Ellis and J.I. Kapusta, Phys. Rev. {\bf C64},
035202.
\bibitem{15a}
A.T. Martell and P.J. Ellis, Phys. Rev. {\bf C69}, 065206 (2004).
\bibitem{16}
R. Rapp and J. Wambach, Eur. Phys. J. {\bf A6}, 415 (1999); Adv. Nucl. Phys.
{\bf 25}, 1 (2000).
\bibitem{17}
H. van Hees and R. Rapp, Phys. Rev. Lett. {\bf 97}, 102301 (2006).
\bibitem{18}
R. Rapp, nucl-th/0701082.
\bibitem{19}
I.M. Dremin, JETP Lett. {\bf 30}, 140 (1979); Sov. J. Nucl. Phys. {\bf 33}, 
726 (1981).
\bibitem{20}
I.M. Dremin, Nucl. Phys. {\bf A767}, 233 (2006); J. Phys. {\bf G35}, 1 (2007);
Int. J. Mod. Phys. {\bf E18}, 1 (2007).
\bibitem{21a}
I.M. Dremin, Nucl. Phys. {\bf A785}, 369 (2007). 
\bibitem{21}
R.P. Feynman, R.B. Leighton and M. Sands, {\it The Feynman Lectures in Physics}
(Addison-Wesley PC Inc., 1963) vol.~1, ch.~31.
\bibitem{22}
M. Goldberger and K. Watson, {\it Collision Theory} (John Wiley and Sons Inc., 1964)
Ch.~11, sect.~3, sect.~4.
\bibitem{23}
A.V. Apanasenko et al., JETP Lett. {\bf 30}, 145 (1979).
\bibitem{26}
S. Kniege and M. Ploskon, nucl-ex/0703008.

\end{thebibliography}
\end{document}